\documentclass[aps,prl,twocolumn,superscriptaddress,notitlepage,nofootinbib,longbibliography]{revtex4-1}
\usepackage{mathrsfs}
\usepackage{epsfig}
\usepackage{graphicx}
\usepackage{amsfonts}
\usepackage[figuresright]{rotating}
\usepackage{amssymb}
\usepackage{amsmath}
\usepackage{dcolumn}
\usepackage{bm}
\usepackage{color}
\usepackage{braket}
\usepackage{float}
\usepackage{units}
\usepackage{xspace}
\begin{document}

\title{ Comment on "Floquet topological phase transition in the $\alpha$--$\mathcal{T}_{3}$ lattice"}

\author{Shujie Cheng}
\address{Department of Physics, Zhejiang Normal University, Jinhua 321004, China}
\date{\today}

\begin{abstract}
A recent paper \cite{Bashab} studied the topological phase transition driven by the circular frequency light in the
$\alpha$--$\mathcal{T}_{3}$ lattice. In this Comment we point out that there is a flaw in
the derivation of the effective Hamiltonian in this paper.
\end{abstract}

\maketitle

In Ref.~\cite{Bashab}, the authors give the Hamiltonian of the static system as
\begin{equation}\label{eq1}
H_{0}(\mathbf{k})=\left(\begin{array}{ccc}
0 & h(\mathbf{k}) \cos \phi & 0 \\
h^{*}(\mathbf{k}) \cos \phi & 0 & h(\mathbf{k}) \sin \phi \\
0 & h^{*}(\mathbf{k}) \sin \phi & 0
\end{array}\right),
\end{equation}
where $\mathbf{k}=\left(k_{x}, k_{y}\right)$ and
$h(\mathbf{k})=\tau\left(e^{i \mathbf{k} \cdot \boldsymbol{\delta}_{1}}+e^{i \mathbf{k} \cdot \boldsymbol{\delta}_{2}}+e^{i \mathbf{k} \cdot \delta_{3}}\right)$
with $\tau$ is the unit of energy,
and present the circular-polarized-light driven Hamiltonian as
\begin{equation}\label{eq2}
H(\mathbf{k}, t)=\left(\begin{array}{ccc}
0 & h(\mathbf{k}, t) \cos \phi & 0 \\
h^{*}(\mathbf{k}, t) \cos \phi & 0 & h(\mathbf{k}, t) \sin \phi \\
0 & h^{*}(\mathbf{k}, t) \sin \phi & 0
\end{array}\right),
\end{equation}
where $\mathbf{A}(t)=A_{0}(\cos \omega t, \sin \omega t)$, where $A_{0}=$ $E_{0} / \omega$ with $E_{0}$ and $\omega$
being the electric field amplitude and the frequency of the radiation, respectively, and
$h(\mathbf{k}, t)=\tau \sum_{j=1}^{3} e^{i(\mathbf{k}+e \mathbf{A}(t) / \hbar) \cdot \delta_{j}}$.

According to the previous works on the circular frequency light driving \cite{circular_light_1, circular_light_2,circular_light_3},
the above Eq.~(\ref{eq1}) is defined in the laboratory frame, and the Eq.~(\ref{eq2}) is defined in the rotating frame. When deriving the
effective Hamiltonian of system, the authors in Ref. \cite{Bashab} takes the high-frequency expansion method, namely,
\begin{equation}\label{eq3}
H_{\mathrm{eff}}(\mathbf{k})=H_{0}(\mathbf{k})+\left[H_{-}(\mathbf{k}), H_{+}(\mathbf{k})\right] / \hbar \omega+O\left(1 / \omega^{2}\right),
\end{equation}
where $H_{\pm}(\mathbf{k})=\frac{1}{T} \int_{0}^{T} d t e^{\mp i \omega t} H(\mathbf{k}, t)$.

Noting that there is a flaw in Eq. (\ref{eq3}), i.e, there appears two different reference frames for one equation at
the same time. As mentioned before, $H_{0}(\mathbf{k})$ in Eq. (\ref{eq3}) is defined in the laboratory frame, whereas $H_{\pm}(\mathbf{k})$ is
derived from $H(\mathbf{k},t)$, i.e, the Hamiltonian in Eq. (\ref{eq2}), which is defined in the rotating frame. This is an incorrect way not supported by
the normal Floquet methods \cite{Floquet_1,Floquet_2,Floquet_3}. One has to select the definite type of reference frame when deriving
the effective Hamiltonian. The correct way to obtain the effective Hamiltonian is to replace the $H_{0}(\mathbf{k})$
with $\frac{1}{T}\int^{T}_{0}dt H(\mathbf{k},t)$, namely
\begin{equation}\label{eq4}
H_{eff}(\mathbf{k})=\frac{1}{T}\int^{T}_{0}dt H(\mathbf{k},t)+\left[H_{-}(\mathbf{k}), H_{+}(\mathbf{k})\right] / \hbar \omega+O\left(1/2\hbar\omega\right).
\end{equation}

\begin{figure}[htp]
\includegraphics[width=0.5\textwidth]{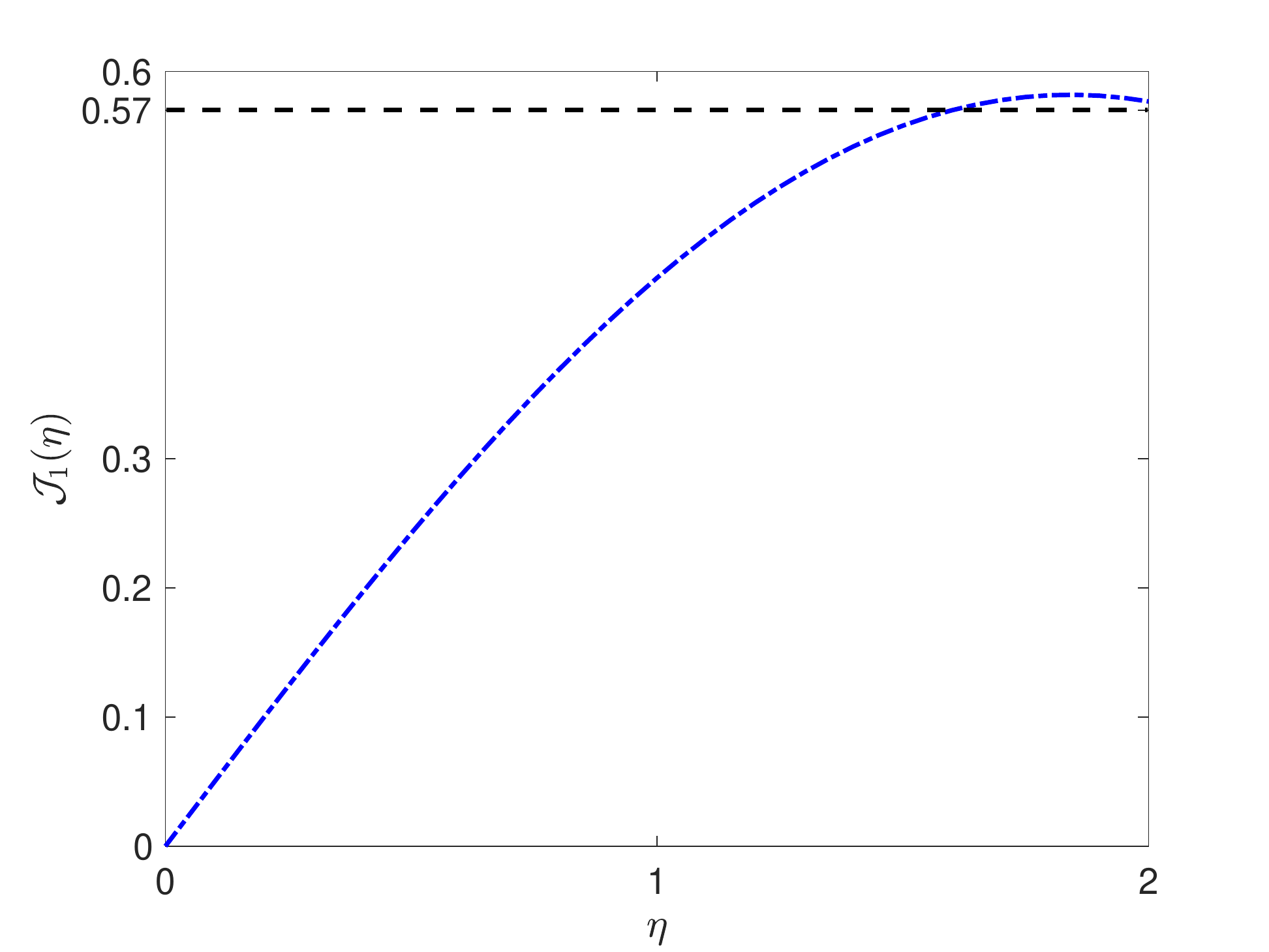}
\caption{The first-order Bessel function $\mathcal{J}_{1}(\eta)$ as a function of the ratio $\eta$.
}
\label{f1}
\end{figure}

In addition, we point out that a rotating frame should be chosen in the derivation of the effective Hamiltonian.
The reason is that when they plotted the energy dispersions (see the Fig. 2 in Ref. \cite{Bashab}), they take
the parameter $\mathcal{J}_{1}(\eta)=0.57$. Here, $\mathcal{J}_{1}(\eta)$ is the first-order Bessel function and $\eta=eE_{0}a/\hbar\omega$
is the ratio of the driving amplitude ($eE_{0}a$) and the photon energy ($\hbar\omega$). We plot the $\mathcal{J}_{1}(\eta)$ as
a function of the ratio $\eta$ in Fig. \ref{f1}. We can see that if $\mathcal{J}_{1}(\eta)$ is equal to 0.57, the chosen $\eta$ is larger than
1, which leads to the divergence of the high-frequency expansion in the laboratory frame (The summation in Eq.~(\ref{eq3}) 
shall contain infinity terms if we transform the matrix $H(\mathbf{k},t)$ into the laboratory frame.) \cite{Floquet_1,Floquet_2,Floquet_3}. 
Thus, we cannot obtain the effective Hamiltonian explicitly in the laboratory frame.

\begin{figure}[htp]
\includegraphics[width=0.5\textwidth]{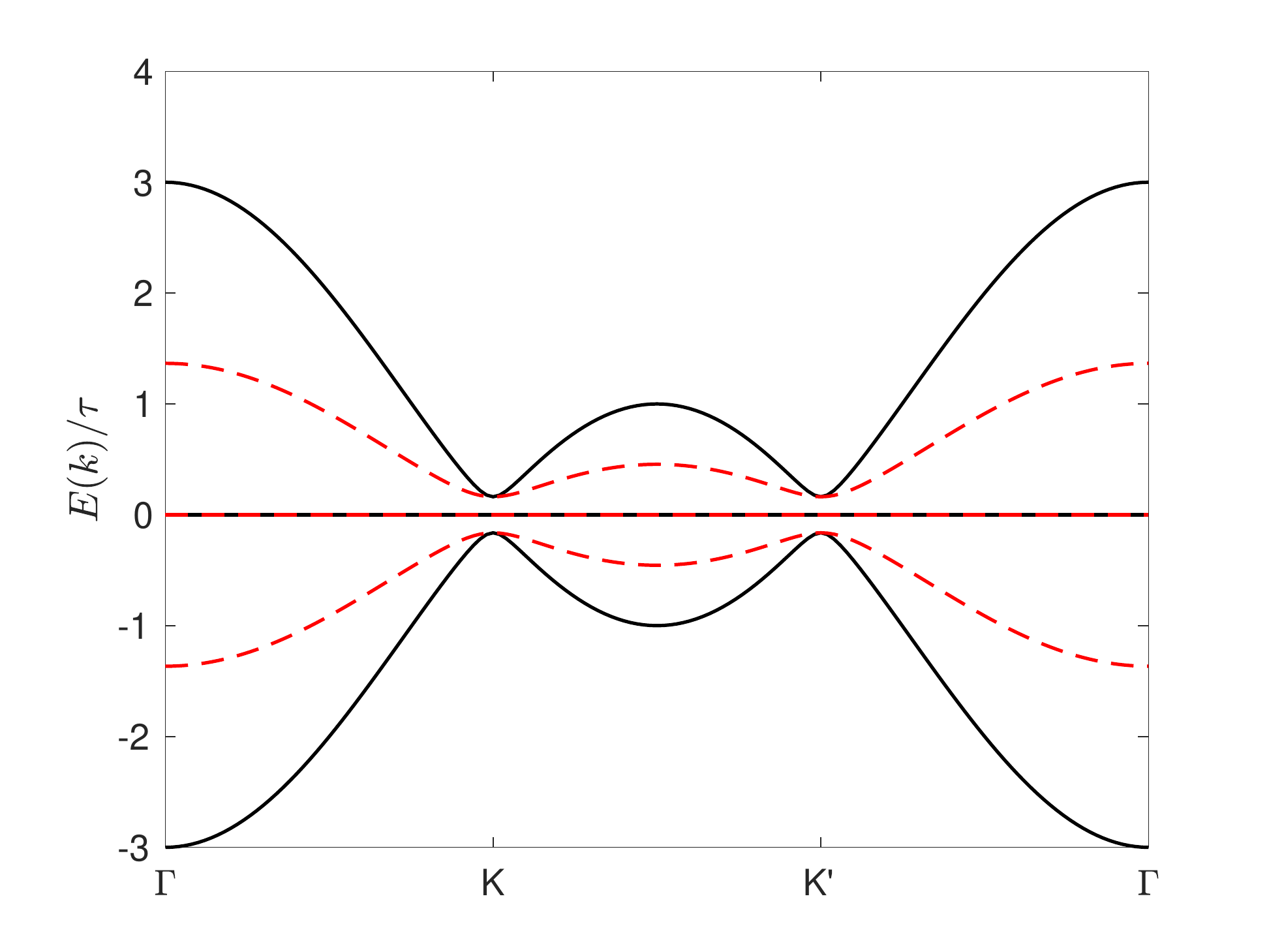}
\caption{The dispersion of the quasi-bands along the high-symmetry path ${\rm \Gamma}$--${\rm K}$--${\rm K'}$--${\rm \Gamma}$.
The black curves are calculated from the incorrect effective Hamiltonian in Ref. \cite{Bashab}, while the red dashed ones are calculated from the
correct effective Hamiltonian in Eq. (\ref{eq5}). Other parameters are $\hbar\omega=9\tau$, $\mathcal{J}_{0}(\eta)=0.4554$,
$\mathcal{J}_{1}(\eta)=0.57$, and $\phi=\pi/4$.}
\label{f2}
\end{figure}

In order to obtain an explicitly effective Hamiltonian, therefore, it is necessary
to select the rotating frame. Within the allowable margin of error, in this way, 
the high-frequency expansion only contains finite terms (see Eq.~(\ref{eq4})) (According to Eq. (\ref{eq3}), 
we can obtain a convergent matrix as well, but it is not the correct effective Hamiltonian).
By the new Eq. (\ref{eq4}), but not the Eq. (\ref{eq3}) in Ref. \cite{Bashab}, we can arrive at the correct effective
Hamiltonian,
\begin{small}
\begin{equation}\label{eq5}
H_{eff }(\mathbf{k})=\left(\begin{array}{ccc}
\gamma(\mathbf{k}) \cos ^{2} \phi & \mathcal{J}_{0}(\eta)h(\mathbf{k}) \cos \phi & 0 \\
\mathcal{J}_{0}(\eta)h^{*}(\mathbf{k}) \cos \phi & -\gamma(\mathbf{k}) \cos 2 \phi & \mathcal{J}_{0}(\eta)h(\mathbf{k}) \sin \phi \\
0 & \mathcal{J}_{0}(\eta)h^{*}(\mathbf{k}) \sin \phi & -\gamma(\mathbf{k}) \sin ^{2} \phi
\end{array}\right).
\end{equation}
\end{small}
Compared to the incorrect effective Hamiltonian in Ref. \cite{Bashab}, in the new Eq. (\ref{eq5}), there appears an extra correction $\mathcal{J}_{0}(\eta)$
on the nearest-neighbor hopping strengths (Here, $\mathcal{J}_0(\eta)$ is the zero-order Bessel function). This correction leads to a significant
deviation between the quasi-bands of the correct effective Hamiltonian and the incorrect one (see details in Fig.~\ref{f2}).

\begin{figure*}[htp]
\centering
\includegraphics[width=\textwidth]{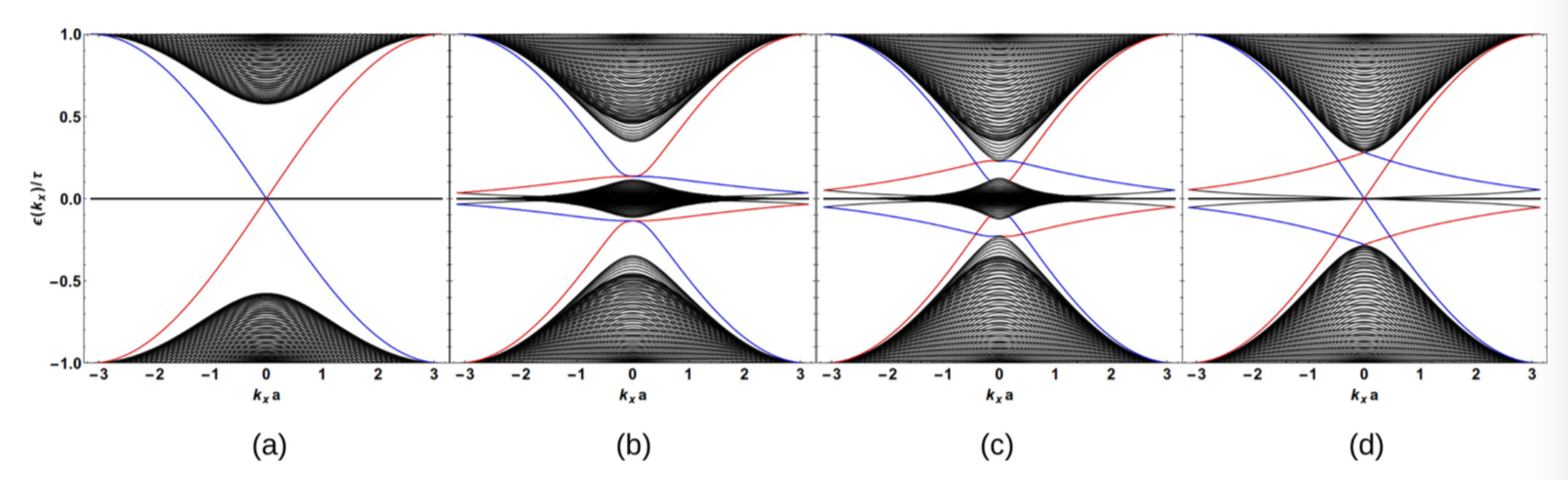}
\centering
\caption{The edge spectrum with armchair edges \cite{Bashab} for (a) $\alpha=0$, (b) $0.5$, (c) $0.8$, and (d) 1.0. 
The red/blue curves represent the edge states, and the black curves represent the bulk bands.}
\label{f3}
\end{figure*}

\begin{figure}[H]
\includegraphics[width=0.5\textwidth]{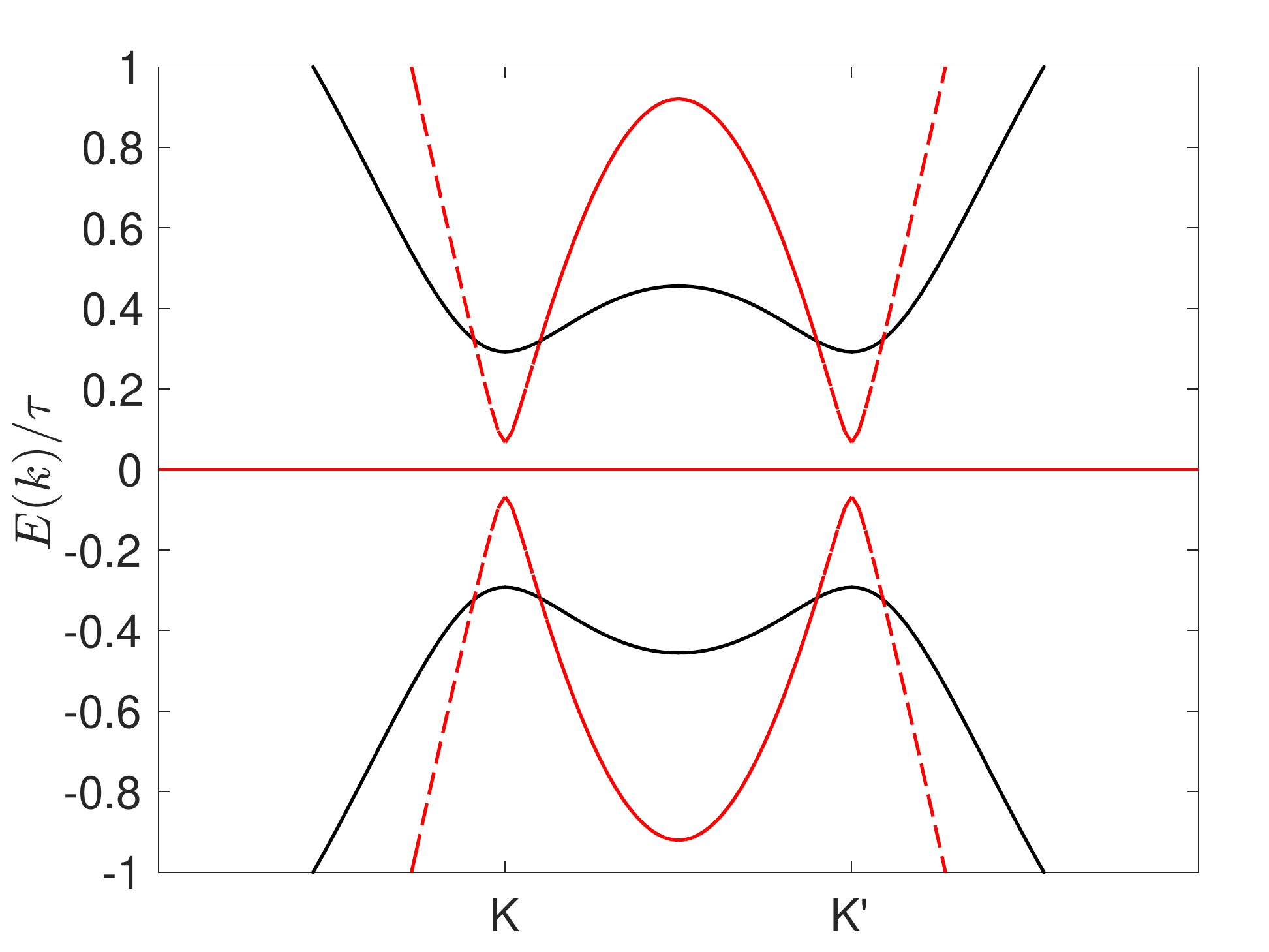}
\centering
\caption{The band dispersions of the correct effective Hamiltonian with $\phi=\pi$ (corresponding to $\alpha=0$) 
The black curves are the dispersions for $\eta=1.6$, and the red dashed curves are the ones for $\eta=0.57$. 
Other parameter is $\hbar\omega=10\tau$.}
\label{f4}
\end{figure}

In addition, an extra incorrect result in the paper \cite{Bashab} is found. In the mentioned paper, the 
authors calculated  the edge-state spectra with armchair edges, as shown in Fig.~\ref{f3} (or Fig. 7 in the mentioned paper). 
From the edge-state spectra in Figs.~\ref{f3}(a)-(d), we can see that there are large band gap where the edge states lie. 
However, here we comment that these results are incorrect. Without loss of generality (for other $\alpha$, the results are 
incorrect and nonphysical, too), we take the cases with $\alpha=0$ as examples. In Fig.~\ref{f3}, the band gap seem to 
be larger than $0.5\tau$. The values are not consistent with the band structures shown in Fig.~\ref{f4}. For $\eta=1.6$, 
the band gaps shall be about $0.3\tau$. For $\eta=0.57$ (In their response, the authors told that $\eta$ shall be $0.57$.), 
the band gaps shall be about $0.1\tau$. In other words, the effect of nontrivial topology of their model can not get correctly 
reflected in the transport measurements through the edge states shown in Fig.~\ref{f3} for the armchair stripe. We carefully 
check the data in their paper used for calculating the spectra, and are surprised to find that they used $\mathcal{J}_{1}(\eta)=0.8$ 
(check it in the mentioned paper: in page 7, right column, the 11-th row from the bottom) in the calculations. As we know, 
no $\eta$ can satisfy $\mathcal{J}_{1}(\eta)=0.8$. Therefore, the results for the chiral edge states are incorrect 
and nonphysical, and even do not mathematically exist.

In conclusion, the results of the dispersion of the quasi-bands, the band gaps, the density-contour 
plots of the Berry curvature, and the Chern numbers, as well as the edge-state spectra presented in Ref. \cite{Bashab} 
are based on the incorrect effective Hamiltonian and parameters, which lead to unbelievable results 
and conclusions. Therefore, we think that our Comment is timely to correct all.


\end{document}